\documentclass[12pt,english]{article}
\usepackage[affil-it]{authblk}
\usepackage{fancyhdr}
\usepackage[top=2.0cm, bottom=2.0cm, left=2.0cm, right=2.0cm]{geometry}
\usepackage{latexsym, amssymb, enumerate, amsmath,amsthm}
\usepackage{color}
\usepackage{xcolor}
\usepackage{hyperref}
\usepackage{tikz}
\usepackage{pgfplots}
\usepackage{pgfplotstable}
\usepackage{changebar}
\usetikzlibrary{decorations.pathreplacing}
\usepackage{tikz}
\usetikzlibrary{decorations}
\usetikzlibrary{decorations.pathmorphing}
\usepackage{url}

    \usepackage{graphicx}
= 0.5\thickmuskip \arraycolsep = 0.3\arraycolsep

\definecolor{dgreen}{rgb}{0,.5,0}
\definecolor{grau}{gray}{.5}
\definecolor{schwarz}{gray}{0}

\newcommand{\Pe}{{\rm Pe}\,}
\newcommand{\bb}[1]{\mathbb{#1}}

\newcommand{\reff}[1]{(\ref{#1})}
\newcommand{\ol}[1]{\overline{#1}}

\newcommand{\av}[1]{\left|#1\right|}

\newcommand{\N}[2]{\left\|#1\right\|_{#2}}

\newcommand{\brkts}[1]{\left(#1\right)}
\newcommand{\ebrkts}[1]{\left[#1\right]}

\newcommand{\brcs}[1]{\left\{#1\right\}}

\newcommand{\pd}[2]{\frac{\partial #1}{\partial #2}}

\newcommand{\bsplitl}[2]{
\begin{equation}
\begin{split}
#1
\end{split}
\label{#2}
\end{equation}}

\newcommand{\csection}[1]{%
  \section[#1]{\centering\Large\MakeUppercase\itshape #1}} 

\DeclareGraphicsExtensions{.png,.eps,.ps}

\begin{document}

\title{Recent advances in the evolution of interfaces:
\\
thermodynamics, upscaling, and universality}
\date{\today{}}

\author{\href{http://www.macs.hw.ac.uk/~ms713}{M. Schmuck}%
	\thanks{\href{mailto:M.Schmuck@hw.ac.uk}{M.Schmuck@hw.ac.uk} (corresponding author)}}
	\affil{\small
	Maxwell Institute for Mathematical Sciences\\
	and School of Mathematical and Computer Sciences\\
	Heriot-Watt University\\
	Edinburgh, EH144AS, UK
	}		
\author{G.A. Pavliotis
	\thanks{\texttt{G.Pavliotis@imperial.ac.uk}}}
	\affil{
	Department of Mathematics\\
	Imperial College London\\
	South Kensington Campus\\
	London SW7 2AZ
	}
\author{S. Kalliadasis%
	\thanks{\texttt{S.Kalliadasis@imperial.ac.uk}}}
	\affil{
	Department of Chemical Engineering\\
	Imperial College London\\
	South Kensington Campus\\
	London SW7 2AZ
	}

\pagestyle{myheadings}
\markright{\emph{M. Schmuck et al.}\hfill Recent advances in the evolution of interfaces \hfill}

\makeatletter
\def\@maketitle{%
  \newpage
  \null
  \vskip 2em%
  \begin{center}%
  \let \footnote \thanks
    {\Large\bfseries \@title \par}%
    \vskip 1.5em%
    {\normalsize
      \lineskip .5em%
      \begin{tabular}[t]{c}%
        \@author
      \end{tabular}\par}%
    \vskip 1em%
    {\normalsize \@date}%
  \end{center}%
  \par
  \vskip 1.5em}
\makeatother

\maketitle

\begin{abstract}
We consider the evolution of interfaces in binary mixtures
permeating strongly heterogeneous systems such as porous media. To this end, we
first review available thermodynamic formulations for binary mixtures based on \emph{general reversible-irreversible couplings} and the associated
mathematical attempts to formulate a \emph{non-equilibrium variational
principle} in which these non-equilibrium couplings can be
identified as minimizers.

Based on this, we investigate two microscopic binary mixture formulations
fully resolving heterogeneous/perforated domains: (a) a flux-driven
immiscible fluid formulation without fluid flow; (b) a momentum-driven
formulation for quasi-static and incompressible velocity fields. In both
cases we state two novel, reliably upscaled equations for binary
mixtures/multiphase fluids in strongly heterogeneous systems by
systematically taking thermodynamic features such as free energies into
account as well as the system's heterogeneity defined on the microscale
such as geometry and materials (e.g. wetting properties). In the context of
(a), we unravel a \emph{universality} with respect to the coarsening rate
due to its independence of the system's heterogeneity, i.e. the well-known
${\cal O}(t^{1/3})$-behaviour for homogeneous systems holds also for
perforated domains.

Finally, the versatility of phase field equations and their
\emph{thermodynamic foundation} relying on free energies, make the
collected recent developments here highly promising for scientific,
engineering and industrial applications for which we provide an example for
lithium batteries.

\vspace{0.25cm}

\end{abstract}

\csection{Introduction}\label{sec:1}
A wide range of problems of scientific, engineering and practical interest
involve the dynamics of interfaces, which by itself is already a delicate
multiscale problem in homogeneous environments. In this article, we make a
further step and increase the number of scales involved by looking at
interfaces evolving in complex heterogeneous systems (CHeSs) such as porous
media, composites and crystals, which are ubiquitous in a wide spectrum of
technological applications. The presence of additional scales brings in a
number of complex phenomena and new effects. For example, wetting of
chemically and/or topographically heterogeneous surfaces, is often
characterised by pinning-depinning effects, steady-state multiplicity and
hysteresis behavior~\cite{Savva2010,Raj2011,Wylock2012,Pradas2016} absent in
wetting of ideally smooth surfaces.

For simplicity, we restrict ourselves here to binary mixtures and look at
extended and systematic non-equilibrium thermodynamic formulations such as
general non-equilibrium reversible-irreversible couplings (GENERIC;
Section~\ref{sec:Gen}). These reversible-irreversible couplings
\cite{Grmela1997,Oettinger1997} are, to the best of our knowledge, currently
the most systematic non-equilibrium thermodynamic formulation available to
reliably describe the dynamics of CHeSs. The necessity of an \emph{extended
theory of thermodynamics} \cite{Mueller1993} is motivated by the paradox of
infinite speed of propagation implied by the classical (parabolic)
formulation not taking a more refined concept of entropy into account. These
new developments became soon also increasingly interesting for mathematicians
to extend the classical least action principle for reversible systems towards
a generalised variational principle, e.g.
\cite{Hackl2008prsa,Hyon2010,Liu2003}. In \cite{Hyon2010,Liu2003}, the
authors first combine the least action principle with gradient flows and
subsequently add a so-called maximum dissipation principle. These
developments together with the elegant theory of optimal transport and
gradient flows based on the Wasserstein distance initiated in
\cite{Jordan1998}, lead to an increased interest to refine the gradient
theory to variationally deduce irreversible/dissipative terms in governing
equations with a continuously increasing list of publications
\cite{Arnrich2012,Hackl2008prsa,Mielke2011}.

Under this non-equilibrium thermodynamic viewpoint, we present recent, systematic upscaling results for binary mixtures in CHeSs by reliably taking a
representative microscopic sub-system into account
\cite{Schmuck2017siap,Schmuck2014aml,Schmuck2012prsa,Schmuck2013nonl}. This
novel formulation (in Section~\ref{sec:EfMaInEv} {\bf (A)} without fluid flow and
in Section~\ref{sec:EfMaInEv} {\bf (B)} for quasi-static flow) serves as a promising tool for multiphase flow in
porous media since it relies on thermodynamic free energies of the
fluid's mixture. Moreover, a first rigorous error quantification has
been derived in \cite{Schmuck2017siap} which we hope to foster
interest in deriving sharp estimates not depending on the classical (and
sub-optimal) truncation argument near the boundary. Hence, this novel upscaled formulation
represents a promising alternative to earlier multiphase flow descriptions
accounting for the permeability of porous media by a Darcy or Brinkman equation for the
momentum but not in the equation for the order parameter governing the interface.
Moreover, a computational investigation of the coarsening process
(Section~\ref{sec:CoRa}) of
binary mixtures for various CHeSs indicates that the classical
rate ${\cal O}(t^{1/3})$ turns out to be \emph{universal} and independent of
perforations \cite{Ververis2017}.

Finally, in Section~\ref{sec:appl}, this
novel upscaled phase field equations allow for a low-dimensional,
effective macroscopic description of lithium intercalation in
composite cathodes of batteries.
The phase field modelling in this context
has been initiated by \cite{Han2004} and since then intensely
investigated in various articles looking at single
particles \cite{Burch2008}
and a consistent thermodynamic description of the Butler-Volmer
reactions \cite{Bazant2013accChemRes}. We present effective
charge transport equations for composite cathodes based on
a binary symmetric electrolyte described by the dilute solution
theory and account for effective
interfacial Butler-Volmer reactions as well as lithium intercalation
in solid crystals which undergo a possible phase separation.

\section{Thermodynamic concepts for modelling binary fluids}\label{sec:2}

We consider binary fluids consisting of species $i\in\brcs{\alpha,\beta}$
with number densities $n_i({\bf x},t)$, total momentum ${\bf M}({\bf x},t)$,
and the mixture's internal energy $u({\bf x},t)$. Hence, the state of the
system can be described by four independent variables ${\bf
z}^*:=\brcs{n_\alpha,n_\beta,{\bf M},u}$. In order to describe a possible
phase transition in binary systems, one generally introduces an order
parameter
\bsplitl{
\phi({\bf x},t)
	:=
	\frac{m_\alpha n_\alpha}{m_\alpha n_\alpha + m_\beta n_\beta}
	\,,
}{phi}
which describes the fraction of a particular species, e.g. $i=\alpha$ here.
Accordingly, this suggests to introduce the total mass density
\bsplitl{
\rho({\bf x},t)
	:=
	m_\alpha n_\alpha
	+ m_\beta n_\beta
	\,,
}{rho}
where $m_i$ denotes the mass of species $i\in\brcs{\alpha,\beta}$. At the
same time, relations \reff{phi} and \reff{rho} can be inverted to
$n_\alpha=\rho \phi/m_\alpha$ and $n_\beta=\rho(1-\phi)/m_\beta$.
Herewith, the system's state ${\bf z}$ is again represented by four independent variables, i.e.,
${\bf z}:=\brcs{\rho,\phi,{\bf M},u}$.

\medskip

Frequently, such binary systems are described as a
\emph{regular solution} \cite{Cahn1958}, which consists of the following
free energy density
\bsplitl{
f_r(\phi)
	:=
	f_s(\phi)
	+f_e(\phi)
	\,,
}{RS}
where
$f_s(\phi):= k_BT \ebrkts{\phi\log\,\phi +(1-\phi)\log\,(1-\phi)}$
is the ideal entropy of mixing and
$f_e(\phi):= z\omega \phi(1-\phi)$
describes the interaction energy between $\alpha$ and $\beta$.
Often, the regular solution free energy \reff{RS} is approximated by
a so-called double-well potential
$w(\phi):=\frac{1}{4a}(1-\phi^2)^2$
allowing for stable numerical schemes.

Finally, we note that the Cahn-Hilliard phase field equation can be derived as a mass
conserving gradient flow of the free energy associated with the regular solution
free energy density $f_r$. It is a simple formulation for first order phase transitions\footnote{
\emph{Ehrenfest's classification scheme \cite{Ehrenfest1933}:}
For temperature $T$, entropy $S$, pressure $p$, and volume $v$, consider the constant Gibbs
free energy $C(T,p):=G=U-TS+pv$ in the $p-T$-plane. First order and second order
phase transitions are then defined as discontinuities (kinks) of first and second
order derivatives of $C$, respectively.
}, e.g.
obtained by quickly quenching a stable single phase solution, i.e., by quickly lowering the
temperature, see \cite{NovickCohen2008} for instance.

\subsection{General non-equilibrium reversible-irreversible couplings}
\label{sec:Gen} Let us first recall the basic building blocks of GENERIC
\cite{Grmela1997,Oettinger2004book,Oettinger1997}, which connect the
essential thermodynamic quantities such as the state vector ${\bf z}$, the
total energy $E({\bf z})$, and the total entropy $S({\bf z})$ by the
following equations
\bsplitl{
	\begin{cases}
\quad
	{\bf z}_t
	=
	{\cal L}({\bf z})\frac{\delta E({\bf z})}{\delta {\bf z}}
	+
	{\cal M}({\bf z})\frac{\delta S({\bf z})}{\delta {\bf z}}
		\,,
\\\qquad
	{\cal L}({\bf z})\frac{\delta S({\bf z})}{\delta {\bf z}}
	=
	0
	\,,
\\\qquad
	{\cal M}({\bf z})\frac{\delta E({\bf z})}{\delta {\bf z}}
	= 0
	\,,
	\end{cases}
}{GENERIC}
where the \emph{antisymmetric matrix}
$
{\cal L}
$
and the \emph{symmetric matrix}
$
{\cal M}
$
are the so-called \emph{Poisson} and \emph{friction} matrices, respectively.
Since the energy $E$ and the Poisson matrix
${\cal L}$
account for the reversible contributions and similarly
the entropy $S$ and the friction matrix $M$ account for the
irreversible elements, the degeneracy
requirements \reff{GENERIC}$_2$--\reff{GENERIC}$_3$
are imposed for cases where reversible and irreversible
quantities are mixed.

For a phase separating binary system described by the state vector ${\bf
z}:=\brcs{\rho,\phi,{\bf M},u}$, the following general total energy and
entropy have been proposed in \cite{Jelic2009PhD}, i.e.,
\bsplitl{
U({\bf z})
	:= \int_D\brkts{
		\frac{{\bf M}^2}{2\rho}
		+u
		+\frac{\lambda_U}{2}\av{\nabla \phi}^2
	}\,d{\bf x}
	\,,
\\
S({\bf z})
	:= \int_D\brkts{
		s(\rho,\phi,u)
		+\frac{\lambda_S}{2}\av{\nabla \phi}^2
	}\,d{\bf x}
	\,,
}{EandS}
where $s$ and $u$ are problem specific entropy and internal energy densities, respectively.
The square gradient penalty, going back to van der Waals \cite{vanDerWaals1892}, is
divided into an energetic contribution and an entropic contribution with the associated
coefficients $\lambda_U$ and $\lambda_S$, respectively. Hence, if one considers
the usual Helmholtz free energy\footnote{Helmholtz free energy describes maximum amount of work at constant volume and temperature}, that means,
$
F({\bf z})= U({\bf z})-k_BT S({\bf z})
	\,,
$
where $U$ is the internal energy, then one can identify the classical regularizing
parameter by $\lambda=\lambda_U-k_BT\lambda_S\,.$

Using \reff{EandS}, the symmetric velocity gradient
${\bf S}:=\nabla {\bf v} +(\nabla {\bf v})^T$, and $\pmb{\tau}=\pmb{\Pi}-p{\bf I}$, where ${\bf I}$ is the
identity matrix and $\pmb{\Pi}$ is the total pressure tensor composed of energetic and
entropic parts, i.e., $\pmb{\Pi}=\pmb{\Pi}_{E}+\pmb{\Pi}_S$,
it has been shown in \cite{Jelic2009PhD} that the
following generalised Cahn-Hilliard based binary mixture formulation,
\bsplitl{
\begin{cases}
\quad
\rho_t
	=
	-{\rm div}\brkts{\rho {\bf v}}
	\,,
	&
\\\quad
\phi_t
	=
	-{\bf v}\cdot\nabla\phi
	+
	{\rm div}\brkts{
		MT\nabla\brkts{
			\frac{1}{T}\brkts{
				\mu^*
				-(\lambda_E-T\lambda_S)\Delta\phi
			}
		}
	}
	\,,
	&
\\\quad
{\bf M}_t
	=
	-{\rm div}\brkts{
		{\bf v}\otimes{\bf M}
	}
	-{\rm div}\brkts{
		\pmb{\Pi}
		+\pmb{\tau}
	}
	\,,
	&
\\\quad
u_t
	=
	- {\rm div}\brkts{u{\bf v}}
	- \frac{1}{2}\pmb{\Pi}_{S}\,:\,{\bf S}
	- \pmb{\tau}\,:\,(\nabla{\bf v})^T
	-{\rm div}\brkts{{\bf j}^q}
	&
\\\quad\qquad
	+\kappa_E\Delta\phi{\rm div}\brkts{
		MT\nabla\brkts{
			\frac{1}{T}\brkts{
				\mu'
				-(\kappa_E-T\kappa_S)\Delta\phi
			}
		}
	}\,,
\end{cases}
}{gCH}
is consistent with the GENERIC framework \reff{GENERIC}. Earlier mathematical studies
guaranteeing thermodynamic dissipation of entropic/irreversible processes in binary mixtures
are \cite{Lowenbrub1998}, for instance.
Around the same time as \cite{Jelic2009PhD}, reduced model formulations, which do not specifically focus on the underlying
reversible-irreversible couplings as \reff{gCH}, have been proposed in \cite{Abels2012},
where a connection to an associated sharp interface description is established,
and in \cite{Liu2003,Hyon2010}, where a generalised varational approach is advocated to
obtain the right reversible-irreversible contributions. Finally, first

\medskip

{\bf System \reff{gCH} fulfils GENERIC:} We can identify the variational
derivatives from \reff{EandS} with the help of Gibbs'\footnote{Gibbs free
energy describes maximum amount of work at constant pressure and temperature}
fundamental equation of thermodynamics \cite[e.g. p. 9]{Oettinger2004book},
here stated in differential form,
\bsplitl{
dU
	=
	-p dV
	+ \mu dN
	+ T dS
	\,,
}{dfGE}
which is a consequence of the first and second law of
thermodynamics applied to the fundamental equilibrium concept of
thermodynamics stating
$
dU
	=
	dW
	+
	dQ
$
for work $dW$ and heat $dQ$.

As intuitively and physically motivated in \cite{Oettinger2004book}, we
introduce the concept of local equilibrium, which amounts to dividing a large
non-equilbrium system with \emph{non-uniform state variables} into small
systems for which we can identify \emph{local state densities}. It is
advantageous to describe non-equilibrium systems of volume $V$ with state
densities such as
\bsplitl{
\rho({\bf x},t)
	:= \frac{m N({\bf x},t)}{V}
	\,,
	\quad
	u({\bf x},t)
	:=\frac{U({\bf x},t)}{V}
	\,,
	\quad\text{and}\quad
	s({\bf x},t)
	:=\frac{S({\bf x},t)}{V}
	\,,
}{dens}
instead of relying on the associated extensive variables
$N$, $U$, $S$ defined with respect to a small (equilibrium)
volume element. Note that the volume $V$ in \reff{dens} can be a
reference volume such as
a small volume element in local thermodynamic equilibrium
or even the total volume of the system of interest.

For variables ${\bf z}^*=\brcs{n_\alpha,n_\beta,{\bf M},u}$ describing binary fluids consisting of species
$i\in\brcs{\alpha,\beta}$ with $n_i:= \frac{N_i}{V}$, we have
$s^*(n_\alpha(\rho,\phi),n_\beta(\rho,\phi),u)=s(\rho(n_\alpha,n_\beta),\phi(n_\alpha,n_\beta),u)$
and \reff{dfGE} reads
\bsplitl{
du
	=
	-p dv
	+\mu_\alpha dn_\alpha
	+\mu_\beta dn_\beta
	+T ds^*
	\,,
}{dfGEbf}
which after integrating over a small volume $v$ in local equilibrium
and the property of constant chemical potentials
$\mu_i={\rm const.}$ in $v$, $i\in\brcs{\alpha,\beta}$, leads to the
following expression for the pressure
\bsplitl{
p
	=
	\mu_\alpha n_\alpha
	+\mu_\beta n_\beta
	+Ts^*
	- u
	\,.
}{p}
Similarly, \reff{dfGEbf} implies the following definitions of
chemical potentials $\mu_i$, $i\in\brcs{\alpha,\beta}$, and
temperature $T$, i.e.,
\bsplitl{
-\frac{\mu_i}{T}
	:=
	\frac{\partial s^*}{\partial n_i}
	\,,
	\quad
	i\in\brcs{\alpha,\beta}
	\,,
\quad\text{and}\quad
\frac{1}{T}
	:= 	
	\frac{\partial s^*}{\partial u}
	\,.
}{myAndT}
Finally, with the relations inverse to \reff{rho} and
\reff{phi}, we get with
$\mu=\mu_\alpha\phi/m_\alpha + \mu_\beta(1-\phi)/m_\beta$
and
$\mu^*=(\mu_\alpha/m_\alpha-\mu_\beta/m_\beta)\rho$ the following
partial derivatives
\bsplitl{
\frac{\partial s(\rho,\phi,u)}{\partial\rho}
	= -\frac{\mu}{T}
	\,,
	\quad
\frac{\partial s(\rho,\phi,u)}{\partial\phi}
	= -\frac{\mu^*}{T}
	\,,
	\quad\text{and}\quad
\frac{\partial s(\rho,\phi,u)}{\partial u}
	= 1/T
	\,.
}{pdOfS}
Thanks to \reff{pdOfS}, the variational derivatives of the total
energy $E$ and entropy $S$ read as stated
in \cite{Jelic2009PhD},
\bsplitl{
\begin{cases}
\quad
\frac{\delta E}{\delta{\bf z}}
	=
	\brcs{-{\bf v}^2({\bf x},t)/2,{\bf v}({\bf x},t),1,-\lambda_E\Delta\phi({\bf x},t)}
	\,,
\\\quad
\frac{\delta S}{\delta{\bf z}}
	=
	\brcs{-\mu({\bf x},t)/T({\bf x},t),{\bf 0},1/T({\bf x},t),-\mu^*({\bf x},t)/T({\bf x},t)-\lambda_S\Delta\phi({\bf x},t)}
	\,.
\end{cases}
}{dEdS}
Also in \cite{Jelic2009PhD}, the following Poisson matrix accounting
for the reversible/convective behaviour has been obtained,
\bsplitl{
\left[{\cal L}({\bf z})\right]({\bf x})
	:=
	-
	\begin{bmatrix}
	0 & \nabla \rho & 0 & 0
	\\
	\rho\nabla & \quad[\nabla{\bf M}+{\bf M}\nabla]^T\quad & \quad u\nabla + {\rm div}\pmb{\Pi}_S\quad & -\nabla\phi
	\\
	0 & \nabla u + \pmb{\Pi}_S\cdot\nabla & 0 & 0
	\\
	0 & \nabla\phi & 0 & 0
	\end{bmatrix}
	\,.
}{L}
%
Above, the decomposition
$\pmb{\Pi}=\pmb{\Pi}_E+\pmb{\Pi}_S$
of the pressure tensor in energetic and entropic contributions
has been applied
such that
\bsplitl{
-{\rm div}\pmb{\Pi}
	=
	-\nabla p
	+\brkts{
		\frac{\delta E_\phi}{\delta\phi}
		-T\frac{\delta S_\phi}{\delta\phi}
	}
	\,,
}{gradPI}
where
\bsplitl{
E_\phi
	:=
	\frac{\lambda_E}{2}\int_D
		|\nabla\phi|^2
	\,d{\bf x}
	\,,
	\quad\text{and}\quad
S_\phi
	:=
	\frac{\lambda_S}{2}\int_D
		|\nabla\phi|^2
	\,d{\bf x}
	\,.
}{EphiSphi}

It leaves to account for irreversible (and additive) contributions such as viscosity,
diffusion, and heat conduction which all enter via the friction matrix ${\cal
M}={\cal M}^H+{\cal M}^D$. With the thermal conductivity $\lambda^q$, the
viscosity $\eta$, the dilatational viscosity $\kappa$, i.e.,
$\hat{\kappa}:=\kappa-\frac{2}{3}\eta$, the symmetric velocity gradient ${\bf
S}:=\nabla{\bf v}+(\nabla {\bf v})^T$, and the components
\bsplitl{
m^H_{{\bf M}{\bf M}}
	& :=
	-\brkts{\nabla\brkts{\eta T}\nabla\,\cdot +\brkts{{\bf I}{\rm div}}\eta T\nabla\,\cdot }^T
	-\nabla\hat{\kappa}T\nabla\,\cdot
	\,,
\\
m^H_{{\bf M}u}
	& :=
	{\rm div}\brkts{\eta T{\bf S}\,\cdot }
	+\nabla\brkts{\frac{\hat{\kappa}}{2}{\rm tr}{\bf S}\,\cdot }
	\,,
\\
m^H_{u{\bf M}}
	& :=
	-\eta T{\bf S}\nabla\,\cdot
	-\frac{\hat{\kappa}T}{2}{\rm tr}{\bf S}\nabla\cdot
	\,,
\\
m^H_{uu}
	& :=
	\frac{\eta T}{2}{\bf S}\,:\,{\bf S}\,\cdot
	+\frac{\hat{\kappa}}{4}({\rm tr}{\bf S})^2
	- {\rm div}\brkts{
		\lambda^qT^2\nabla\,\cdot
	}
	\,,
}{M^Hels}
as defined in \cite{Jelic2009PhD,Oettinger2004book}, the general friction matrix from hydrodynamics
reads as follows
\bsplitl{
{\cal M}^H
	:=
	\begin{bmatrix}
	0 & 0 & 0 & 0
\\
	0 & m_{{\bf M}{\bf M}}^H & m_{{\bf M}u}^H & 0
\\
	0 & m_{u{\bf M}}^H & m_{uu}^H & 0
\\
	0 & 0 & 0 & 0
	\end{bmatrix}
	\,,
}{M^H}
and the diffusive contribution \cite{Jelic2009PhD} follows due to the
symmetry and degeneracy requirements, i.e.,
\bsplitl{
{\cal M}^D
	:=
	\begin{bmatrix}
	0 & 0 & 0 & 0
	\\
	0 & 0 & 0 & 0
	\\
	0 & 0 & m_{uu}^D & m_{u\phi}^D
	\\
	0 & 0 & m_{\phi u}^D & m_{\phi\phi}^D
	\end{bmatrix}
	\,,
}{M^D}
where
\bsplitl{
m_{uu}^D
	& :=
	-\kappa_E\Delta\phi{\rm div}\brkts{
		MT\nabla\brkts{\kappa_E\Delta\phi\,\cdot}
	}
	\,,
\\
m_{u\phi}^D
	& :=
	-\kappa_E\Delta\phi{\rm div}\brkts{
		MT\nabla\,\cdot
	}
	\,,
\\
m_{\phi u}^D
	& :=
	-{\rm div}\brkts{
		MT\nabla\brkts{
			\kappa_E\Delta\phi\,\cdot
		}
	}
\\
m_{\phi\phi}^D
	& :=
	-{\rm div}\brkts{MT\nabla\,\cdot}
	\,.
}{M^Dels}
%

\subsection{Variational approaches for irreversible systems: gradient flows, least action, and maximum dissipation principles}
\label{sec:GeCaVa}

Motivated by the generalisation of classical mechanics towards dissipative
effects by a so-called dissipation potential, e.g. as explained in
\cite{Leech1965book}, it seems to become increasingly popular to combine
such a dissipation potential concept with a gradient flow
of the free energy associated with the non-equilibrium system of interest.
For instance, in \cite{Hackl2008prsa}, a variational formulation has been
developed based on a maximum dissipation principle which can be related
to a minimum principle for a dissipation potential. Here, we briefly
motivate these ideas related to binary mixtures as discussed in
\cite{Hyon2010}, extending the earlier work on a least action principle
combined with a gradient flow \cite{Liu2003}. The authors in \cite{Hyon2010} propose
variational principles to derive the following evolution equations taking
thermodynamic principles such as reversible (least action) and
irreversible processes (maximum dissipation)
into account, i.e.,
\bsplitl{
\begin{cases}
\quad
{\bf v}_t
	+{\bf v}\cdot\nabla{\bf v}
	-\eta\Delta{\bf v}
	+\nabla p
	=
	\lambda f
	\,,&
\\\quad\qquad
{\rm div}{\bf v}
	= 0
	\,,
\\\quad
\phi_t
	+{\bf v}\cdot\nabla\phi
	=
	-\gamma\Delta\brkts{
		\Delta\phi
		-w(\phi)
	}
	\,,
\end{cases}
}{ImFl}
where $\eta$ is the viscosity, $\gamma$ denotes an elastic relaxation time of the system, and
$\lambda$ corresponds to the surface tension and
$
f
	:=
	{\rm div}\brkts{
		\nabla\phi\otimes\nabla\phi
		-f_{\rm mix}(\phi,\nabla\phi){\bf I}
	}
$
is the associated force. Moreover, the regular solution character of the
immiscible fluid is approximated by the classical double-well potential $w(\phi)=1/4(1-\phi^2)^2$.
We note that related and modified systems
have been proposed in \cite{Abels2012} by solely relying on local and global dissipation
inequalities and frame indifference.
In order to systematically motivate \reff{ImFl}, the authors in \cite{Hyon2010} combine the classical
least action principle (LAP) for reversible processes with a
maximum dissipation principle (MDP) for irreversible contributions
towards a so-called energetic variational approach (EVA)
for complex fluids.

By identifying the kinetic energy
$
E_{\rm kin}({\bf v})
	:=
	\int_{D}\brkts{
		\frac{\rho}{2}{\bf v}^2
	}\,d{\bf x}
$
and the elastic mixing energy associated with the
Cahn-Hilliard equation, i.e.,
\bsplitl{
E_{\rm mix}(\phi)
	:=
	\int_D\brkts{
		f_{\rm mix}(\phi,\nabla\phi)
	}\,d{\bf x}
	\,,
}{E_mix}
for
$
f_{\rm mix}(\phi,\nabla\phi)
		:=\lambda w(\phi)
		+\frac{\lambda}{2}|\nabla\phi|^2
	\,,
$
one can define the following total energy
\bsplitl{
E_{\rm tot}({\bf v},\phi)
	:=
	E_{\rm kin}({\bf v})
	+
	E_{\rm mix}(\phi)
	\,,
}{E_t}
which includes reversible and irreversible processes governing the
evolution of binary immiscible and incompressible fluids.
The above mentioned least action principle is based on the following
flow map \cite{Liu2003},
\bsplitl{
\begin{cases}
\quad
{\bf x}_t({\bf X},t)
	=
	{\bf v}({\bf x}({\bf X},t),t)
	\,,
\\\quad
{\bf x}({\bf X},0)
	= {\bf X}
	\,,
\end{cases}
}{FM}
which maps the so-called Lagrangian material coordinate ${\bf X}$
into the Eulerian coordinate ${\bf x}({\bf X},t)$. As proposed in
\cite{Liu2003}, we can rewrite the total energy \reff{E_t} in Lagrangian coordinates as the following action functional
\footnote{We note that this generalisation from the total energy \reff{E_t}
to the action functional \reff{Act} can be motivated from
related concepts in classical mechanics where the total energy
represents the Hamiltonian $H=K+V$ whereas the
Lagrangian is defined by $L:= K - V$ with $K$ and $V$ for the kinetic
and potential energy, respectively. In
\cite{Leech1965book} for instance, a related generalisation of classical
mechanics to account for dissipation, e.g. by Rayleigh's dissipation
function $R=\frac{1}{2}\sum k_i\dot{x}_i^2$ is formulated with
the associated generalised Lagrange equation
$
\frac{d}{dt}\brkts{
		\frac{\partial L}{\partial\dot{x}_i}
	}
	-\frac{\partial L}{\partial\dot{x}_i}
	+\frac{\partial R}{\partial\dot{x}_i}
	=
	0
	\,.
$
}
\bsplitl{
A({\bf x})
	=
	\int_0^T\int_{D}\brkts{
		\frac{\rho}{2}\av{{\bf x}_t}^2
		-\lambda w(\phi({\bf x}({\bf X},t),t))
		-\frac{\lambda}{2}\av{\nabla_{\bf x}\phi({\bf x}({\bf X},t),t)}^2
	}\,d{\bf X}
	dt
	\,.
}{Act}
Computing the the variation with respect to the kinetic energy gives the Euler equation
\bsplitl{
{\bf v}_t
	+{\bf v}\cdot\nabla{\bf v}
	+\nabla \hat{p}
	=
	-\lambda{\rm div}\brkts{
		\nabla\phi\otimes\nabla\phi
		-f_{\rm mix}(\phi,\nabla\phi){\bf I}
	}\,,
}{Euler}
where the right-hand side appears due to the elastic mixing energy.
The pressure $p$ plays the role of a Lagrange multiplier in \reff{Euler},
if we additionally impose the incompressibility constraint ${\rm div}{\bf v}=0$.

In order to account for the dissipative part in \reff{Euler}, we introduce the
dissipation potential $\Upsilon:=\frac{1}{2}\int_D\eta\av{\nabla {\bf v}}^2\,d{\bf x}$
following the MDP advocated in \cite{Hyon2010}. Hence, maximizing dissipation by
imposing $\frac{\delta\Upsilon}{\delta {\bf v}}=0$ leads to the incompressible Stokes
equation with Lagrange multiplier $\tilde{p}$. Herewith, the Euler equation turns into
the following incompressible momentum equation
\bsplitl{
{\bf v}_t
	+{\bf v}\cdot\nabla{\bf v}
	-\eta\Delta{\bf v}
	+\nabla p
	=
	-\lambda{\rm div}\brkts{
		\nabla\phi\otimes\nabla\phi
		-f_{\rm mix}(\phi,\nabla\phi){\bf I}
	}\,,
}{IME}
where the pressure gradient follows from $\nabla p=\nabla\hat{p}+\nabla\tilde{p}$.
Finally, minimizing the mixing energy
\bsplitl{
E_{\rm mix}(\phi)
	:=
	\int_0^T\int_{D}\brkts{
		\lambda w(\phi({\bf x}({\bf X},t),t))
		+\frac{\lambda}{2}\av{\nabla_{\bf x}\phi({\bf x}({\bf X},t),t)}^2
	}\,d{\bf X}
	dt
	\,,
}{E_mix}
stated in Lagrangian coordinates in the form of a continuous limit of a gradient descent
leads to the phase field equation under the following (mass-conserving/$H^{-1}$) gradient flow
\bsplitl{
\phi_t
	+{\bf v}\cdot\nabla\phi
	=
	-\frac{\gamma}{\lambda}\frac{\delta E_{\rm mix}}{\delta\phi}
	=
	\gamma\brkts{
		\Delta\phi
		-w'(\phi)
	}
	\,.
}{CH}
The GENERIC framework (Section~\ref{sec:Gen}) and the concept of optimal
transport based on entropy to define gradient flows \cite{Jordan1998} has led
to an increased interest in a rigorous formulation of a general variational
principle for reversible-irreversible couplings, e.g. the so-called dual
dissipation potential concept \cite{Mielke2011} and the related work
\cite{Arnrich2012}. However, a general variational principle allowing to
arrive at GENERIC by minimising functionals that can be systematically
indentified for the underlying physical problem seems still not to be
available at this time, to the best of our knowledge, and represents an
interesting open problem.

\section{Interfacial dynamics in heterogeneous systems}
\label{sec:IntDyn}

Phase field equations represent a convenient computational formulation to numerically
study the evolution of interfaces arising in phase separated mixtures. In the context
of multiphase flow in porous media, there exist various effective macroscopic formulations
such as the generalized Darcy law \cite{Leverett1941,Muskat1936},
\bsplitl{
{\bf v}_i
	=
	-\hat{\kappa}_{i}^r\frac{\hat{\kappa}}{\mu_i}\brkts{
		\nabla p_i
		-f_i
	}
	\,,
	\text{for }i\in\brcs{w,o}
	\,,
}{gDL}
where
$w$ stands for water, $o$ for oil, $\hat{\kappa}_i^r$ the relative
permeability tensor of phase $i$, $\hat{\kappa}$ the absolute permeability tensor,
$\mu_i$ the dynamic viscosity of phase $i$, $p_i$ the pressure of phase
$i$, and $f_i$ a external force such as gravitation. Following this strategy, the authors
of \cite{Bosia2014arXiv} study a system where they combine the Cahn-Hilliard equation
with the Brinkman equation.

Here, we would like to advocate a recently proposed alternative description of interfacial transport of mixtures in highly heterogeneous systems such
as porous media. The important novelty is the generally neglected upscaling of the equation
governing the order parameter, i.e., the evolution of the interface.
The key novelties are the a rigorous
\cite{Schmuck2017siap,Schmuck2013nonl}
and systematic derivation of effective macroscopic phase field equations
\cite{Schmuck2012prsa,Schmuck2014aml}
by reliably taking the pore geometry into account as well as
the thermodynamic nature of the mixture by its specific free energy
density
such as $f_{\rm mix}$ in EVA
or
the entropy density $s(\rho,\phi,u)$ in GENERIC.
For simplicity, we consider the following two scenarios: {\bf (A)} interfacial transport without fluid flow/momentum transport; and {\bf (B)} interfacial transport under quasi-static flow.

\medskip

{\bf (A) Interfacial transport without fluid flow/momentum transport.} In the
following, we will describe the evolution of the interface of an immiscible
fluid forming a liquid/liquid or a liquid/gas interface. Additionally, we
would like to account for the so-called contact angle formed between
fluid/fluid interface and a solid surface. A contact angle of 90$^\circ$ is
referred to as neutral wetting which amounts to a material wetting property
$a=0$, whereas hydrophobic and hydrophilic materials are characterised by
$a<0$ and $a>0$, respectively. Herewith, we are able to account for contact
angles in the phase field formulation by the following inhomogeneous Neumann
boundary condition \cite{Ververis2017,Wylock2012}
\bsplitl{
{\bf n}\cdot\nabla\phi
	=
	g\,,
}{WBC}
where $g({\bf x}):=-\frac{\gamma}{C_h}a({\bf x})$ and
$C_h$ is the Cahn number $\lambda/L$ for a characteristic length scale
$L$ and $\gamma=2\sqrt{2}\phi_e/3\sigma_{lg}$. The variable $\phi_e$ denotes
a local equilibrium limiting value of the free energy $E_{\rm mix}$
and $\sigma_{lg}$ the liquid-gas surface tension.
\begin{figure}
\center
\includegraphics[width=8.5cm]{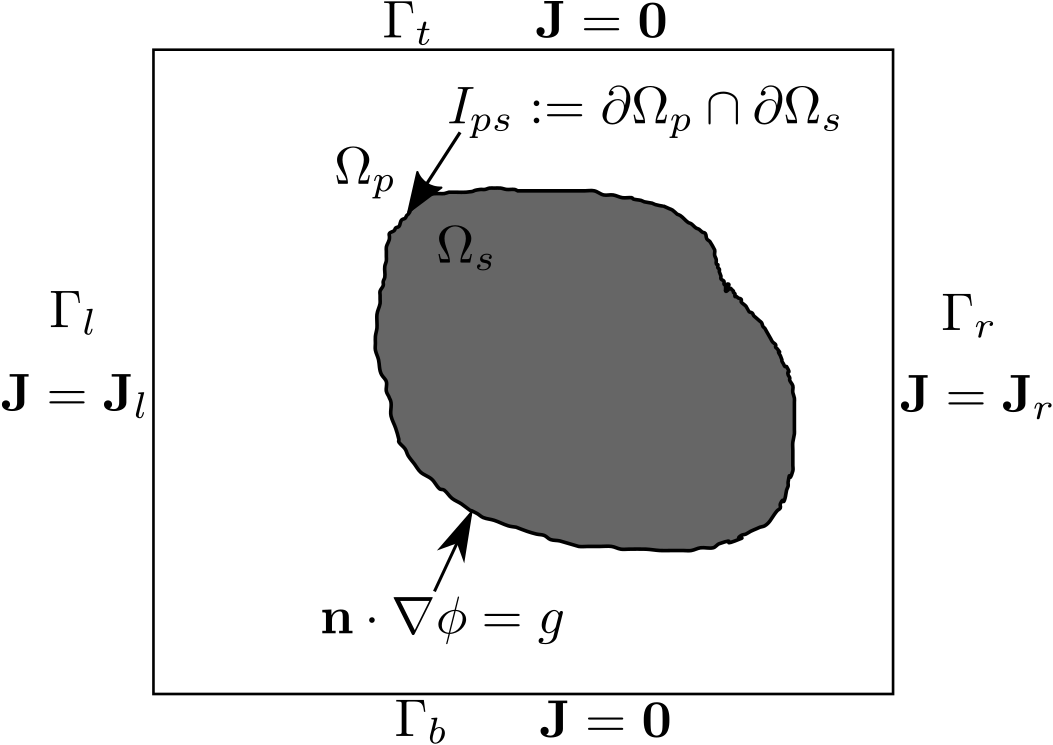}
\caption{Flux-driven interfacial evolution taking into
account specific wetting properties of an obstacle $\Omega_s$ by the
inhomogeneous Neumann boundary condition $g=-\frac{\gamma}{C_h}a({\bf x})$
on $I_{ps}:=\partial\Omega_p\cap\partial\Omega_s$.}
\label{fig:Dom}
\end{figure}

Hence, neglecting the momentum transport in thermodynamically motivated
phase field formulations from \reff{sec:Gen} and \reff{sec:GeCaVa}, i.e.,
\reff{gCH}$_2$ and \reff{ImFl}$_3$, respectively, leads to the following
interfacial evolution problem
\bsplitl{
\pmb{{\bf [A]}}\qquad
\begin{cases}
\quad
\phi_t
	=
	{\rm div}\brkts{
		M\nabla\brkts{
			\frac{1}{\lambda}w'(\phi)
			-\lambda\Delta\phi
		}
	}
	\,,
	&\quad\text{in }\Omega_p
	\,,
\\\qquad
M\nabla\brkts{
			\frac{1}{\lambda}w'(\phi)
			-\lambda\Delta\phi
	}\cdot{\bf n}
	= J_l:={\bf n}\cdot{\bf J}_l
	\,,
	&\quad\text{on }\Gamma_l
	\,,
\\\qquad
M\nabla\brkts{
			\frac{1}{\lambda}w'(\phi)
			-\lambda\Delta\phi
	}\cdot{\bf n}
	= J_r:={\bf n}\cdot{\bf J}_r
	\,,
	&\quad\text{on }\Gamma_r
	\,,
\\\qquad
{\bf n}\cdot\nabla\phi
	= g({\bf x})
	\,,
	&\quad\text{on }I_{ps}
	\,,
\\\qquad
{\bf n}\cdot\nabla\phi
	= 0
	\,,
	&\quad\text{on }\Gamma_t\cup\Gamma_b
	\,,
\\\qquad
{\bf n}\cdot\nabla\Delta\phi
	= 0
	\,,
	&\quad\text{on }I_{ps}\cup\Gamma_t\cup\Gamma_b
	\,,
\\\qquad
\phi({\bf x},0)
	= \phi_0({\bf x})
	&\quad\text{in }\Omega_p
	\,,
\end{cases}
}{IEP}
where $J_l$ and $J_r$ are fluxes imposed such that they
drive the interface from the left to the right while
neglecting momentum transport for simplicity. For the definition of the
variables describing the domain $\Omega=\Omega_p\cup\Omega_s$ and
its boundary $\partial\Omega=\Gamma_l\cup\Gamma_r\cup\Gamma_t\cup\Gamma_b$
as well as its interfaces $I_{ps}:=\partial\Omega_p\cap\partial\Omega_s$,
we refer to Fig.~\ref{fig:Dom}.
\begin{figure}
\center
\includegraphics[width=6.0cm]{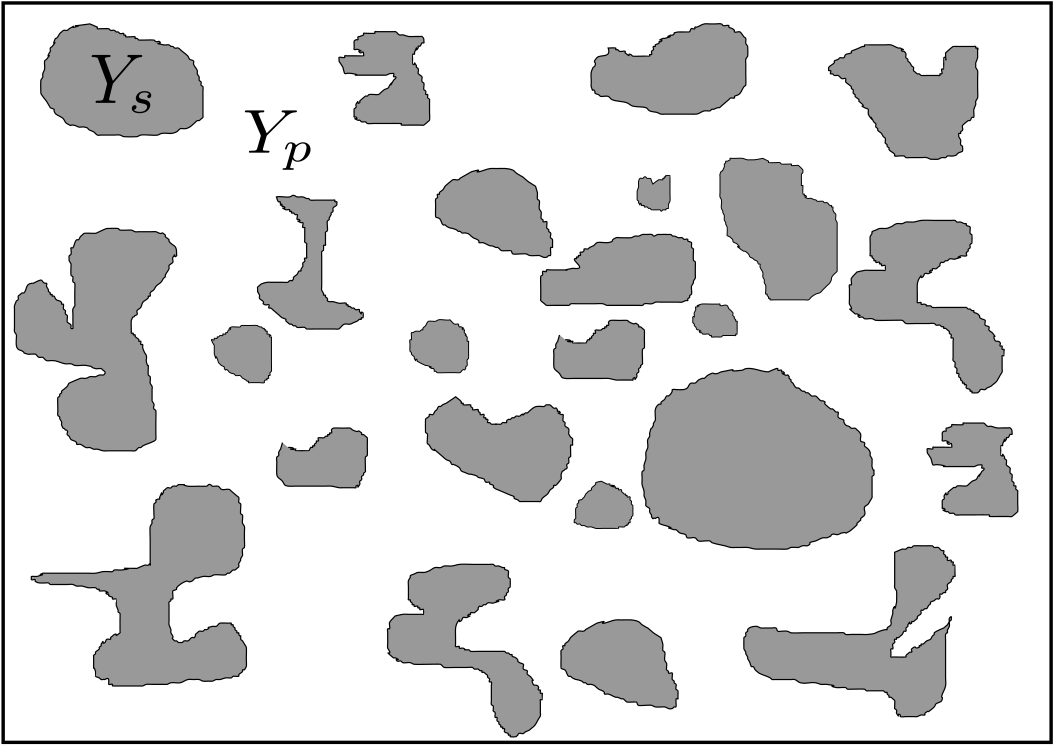}
\hspace{0.5cm}
\includegraphics[width=8.5cm]{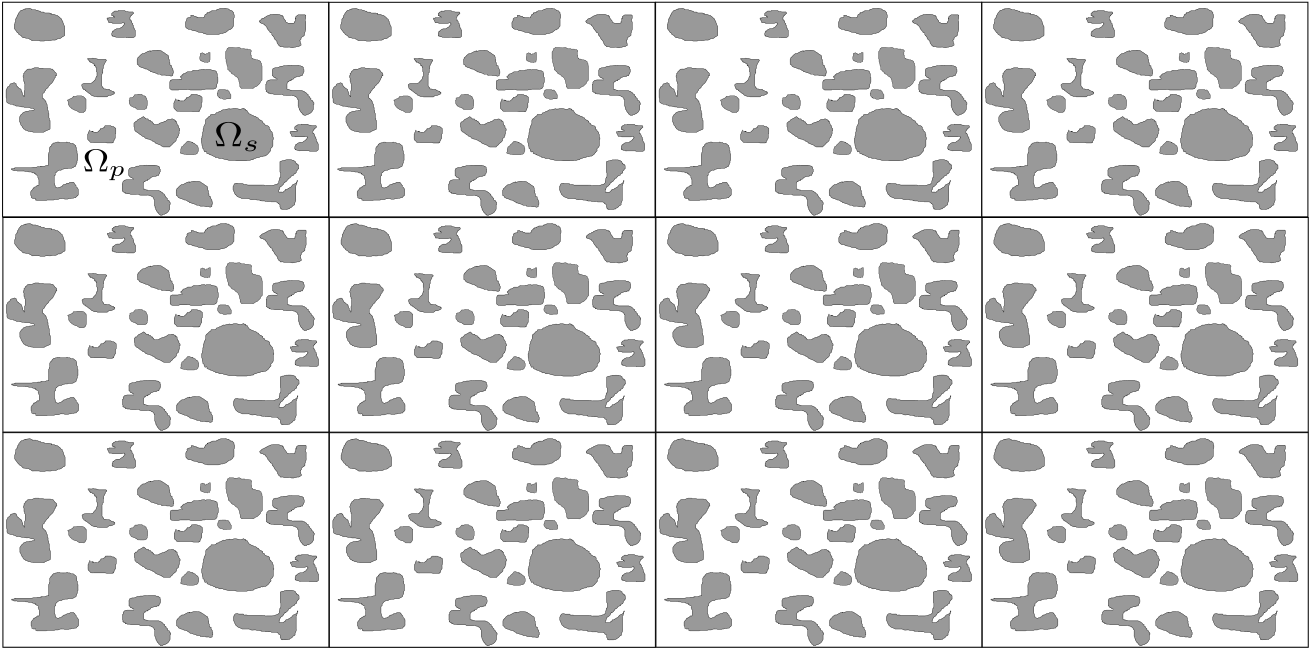}
\caption{The upscaling of CHeSs relies on the ability to identify a
characteristic reference sub-system $Y$. {\bf Left:} Periodic reference cell $Y:=Y_p\cup Y_s$ with representative pore space $Y_p$ and characteristic solid phase $Y_s$.
{\bf Right:} Periodic porous medium $\Omega=\Omega_p\cup\Omega_s$ obtained as a covering of cells $Y$.}
\label{fig:refCell}
\end{figure}
We would like to note that the perforated domain $\Omega_p$
can be defined as the subset of a porous medium $\Omega=\Omega_p\cup\Omega_s$
which is defined as the periodic covering of a reference cell $Y$, see Fig.~\ref{fig:refCell}, but restricted to the pore space $Y_p$.
Herewith, a so-called \emph{heterogeneity parameter}
$\epsilon:=\frac{\ell}{L}$ characterising the porous medium is
systematically defined as the quotient of the length of the representative
porous cell $Y$ divided by the macroscopic length $L$ of the porous
medium of interest. If one looks for solutions of \reff{IEP} in a
perforated domain $\Omega_p$ by such a periodic covering, then one
can generally find $\epsilon$-dependent microscopic formulations, i.e.,
\reff{IEP} rewritten by substituting $\phi$ with $\phi^\epsilon$,
$\Omega_p$ with $\Omega_p^\epsilon$, and $I_{ps}$ with $I_{ps}^\epsilon$.
For notational convenience, we do not explicitly state such an
$\epsilon$-dependence of the microscopic problem here except where
it is necessary for the sake of clarity.

\begin{figure}
\center
\includegraphics[width=5.5cm]{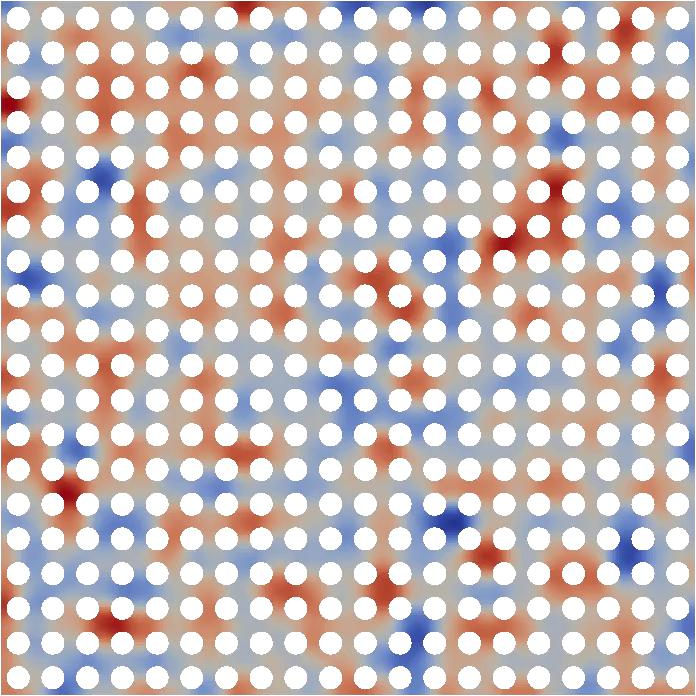}
\hspace{0.5cm}
\includegraphics[width=5.5cm]{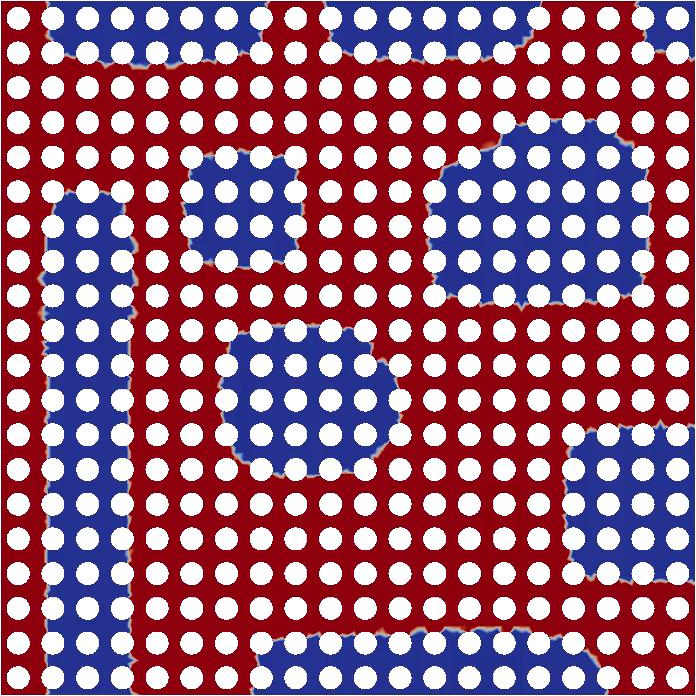}
\\[0.5cm]
\includegraphics[width=5.5cm]{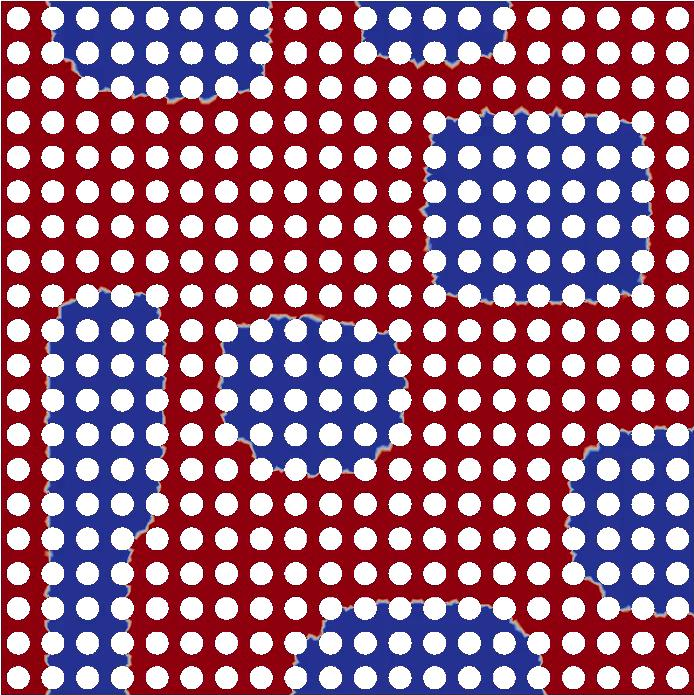}
\hspace{0.5cm}
\includegraphics[width=5.5cm]{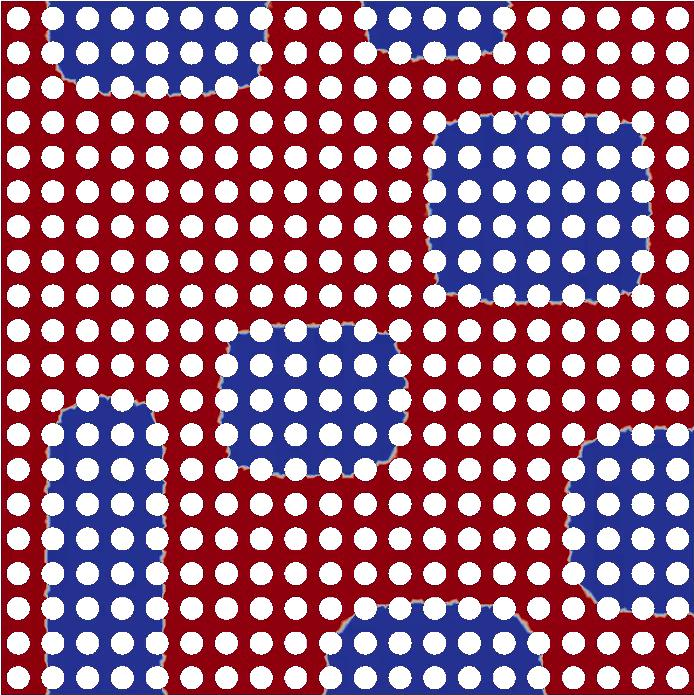}
\caption{Time evolution [time steps 1 (top left), 5 (top right), 10 (bottom left),
and 20 (bottom right)] under critical random initial conditions
$\phi_0({\bf x})=0.5+\xi({\bf x}_h)$ with $\xi({\bf x}_h)\in{\cal N}(0.5,1/8)$,
${\bf x}_h\in{\cal T}_h$, where ${\cal T}_h$ denotes the triangulation of the
connected pore space $\Omega_p$ obtained by removing periodically placed disks from the unit square. Hence, the computations are based on the double-well
potential $w(\phi)=100\phi^2(1-\phi)^2$.
}
\label{fig:mEv}
\end{figure}

\medskip

{\bf (B) Interfacial transport under quasi-static flow.}
We want to generalize {\bf (A)} towards fluid flow. To this end, we consider a horizontal,
quasi-static flow field defined in a periodic reference cell, see Fig.~\ref{fig:refCell},
and driven by a constant, horizontal driving force ${\bf e}_1$, where ${\bf e}_1$ is the
canonical Euclidean basis.
Hence, we define the fluid velocity to be the solution of the following
\emph{periodic cell problem}
\bsplitl{
\pmb{{\bf [B1]}}\qquad
\begin{cases}
\quad
-\mu\Delta_{\bf y} {\bf v}
	+\nabla_{\bf y} p
	= {\bf e}_1
	\,,
	&\quad\text{in }Y_p
	\,,
\\\quad
{\rm div}_{\bf y}{\bf v}
	= 0
	\,,
	&\quad\text{in }Y_p
	\,,
\\\quad
{\bf v}
	= {\bf 0}
	\,,
	&\quad\text{on }I^Y_{ps}:=\partial Y_p\cap\partial Y_s
	\,.
\end{cases}
}{perCP}
For large P\'eclet numbers scaling inversely proportional with heterogeneity, i.e., $\Pe = \Pe_0/\epsilon$, a periodic wetting characterization $g({\bf y})=g({\bf x}/\epsilon)$ of the
porous medium, and the periodic fluid velocity
${\bf v}({\bf y})={\bf v}({\bf x}/\epsilon)$,
we can write the \emph{microscopic interfacial evolution problem}
as follows
\bsplitl{
\pmb{[{\bf B2}]}\qquad
\begin{cases}
\quad
\phi_t
	=
	-\frac{\Pe_0}{\epsilon}{\bf v}\cdot\nabla\phi
	+{\rm div}\brkts{
		\nabla \frac{1}{\lambda}w'(\phi)
		-\lambda\Delta\phi
	}
	\,,
	&\quad\text{in }\Omega_p
	\,,
\\\qquad
\ebrkts{-\frac{\Pe_0}{\epsilon}{\bf v}\phi
	+\nabla\brkts{
			\frac{1}{\lambda}w'(\phi)
			-\lambda\Delta\phi
	}
	}\cdot{\bf n}
	= J_l:={\bf n}\cdot{\bf J}_l
	\,,
	&\quad\text{on }\Gamma_l
	\,,
\\\qquad
\ebrkts{-\frac{\Pe_0}{\epsilon}{\bf v}\phi
	+\nabla\brkts{
			\frac{1}{\lambda}w'(\phi)
			-\lambda\Delta\phi
	}
	}\cdot{\bf n}
	= J_r:={\bf n}\cdot{\bf J}_r
	\,,
	&\quad\text{on }\Gamma_r
	\,,
\\\qquad
{\bf n}\cdot\nabla\phi
	= g({\bf x}/\epsilon)
	\,,
	&\quad\text{on }I_{ps}
	\,,
\\\qquad
{\bf n}\cdot\nabla\phi
	= 0
	\,,
	&\quad\text{on }\Gamma_t\cup\Gamma_b
	\,,
\\\qquad
{\bf n}\cdot\nabla\Delta\phi
	= 0
	\,,
	&\quad\text{on }I_{ps}\cup\Gamma_t\cup\Gamma_b
	\,,
\\\qquad
\phi({\bf x},0)
	= \phi_0({\bf x})
	\,,
	&\quad\text{in }\Omega_p
	\,,
\end{cases}
}{MIEP}
where we have set the mobility to $M=1$ for simplicity.

\begin{figure}
\center
\includegraphics[width=3.95cm]{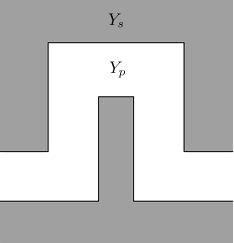}
\caption{Reference sub-substem $Y:=Y_p\cup Y_s$ defining the representative
microscale.}
\label{fig:pY}
\end{figure}

\begin{figure}
\center
\includegraphics[width=6.2cm]{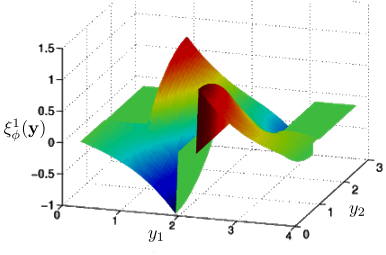}
\hspace{1cm}
\includegraphics[width=6.2cm]{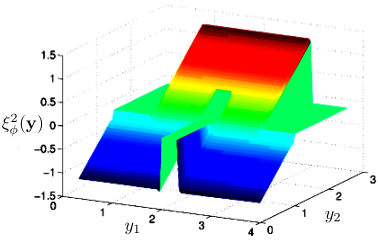}
\\
\includegraphics[width=6.2cm]{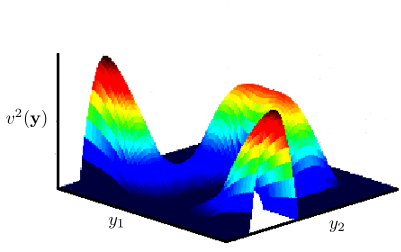}
\hspace{1cm}
\includegraphics[width=6.2cm]{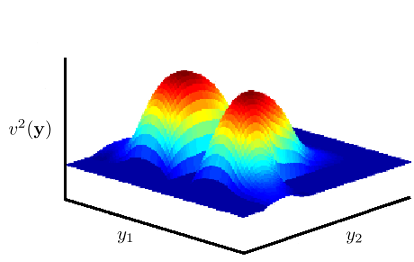}
\caption{{\bf Top line:} Heterogeneous system correctors $\xi^1$ and $\xi^2$ solving the
cell problem (reference sub-system) \reff{ellY} and the perturbed straight channel Fig.~\ref{fig:pY}. {\bf Bottom line:} Components
$v^1$ and $v^2$ solving \reff{perCP} in the reference cell Fig.~\ref{fig:pY}/}
\label{fig:XiFl}
\end{figure}

\subsection{Effective macroscopic interfacial evolution and error quantification}
\label{sec:EfMaInEv} The microscopic formulations \reff{IEP} and
\reff{perCP}-\reff{MIEP} lead to computationally high-dimensional problems
since the mesh size needs to be chosen much smaller than the heterogeneity
$\epsilon$. Also defining the pore and solid space together with the
associated interfaces, which are generally obtained with the help of imaging
tools, is rather challenging for such complex geometries such as porous media. Moreover, the subsequent mesh generation is also more
demanding due to the complex geometries requiring a large number of degrees
of freedom for a reliable resolution.

As a consequence, one can accelerate the computation of practical problems by
identifying the characteristic pore geometry for a smaller representative
volume element, e.g., by a reference cell $Y$ as depicted in
Fig.~\ref{fig:refCell}, which contains all the relevant information about
geometry. For such a reference cell, the mesh generation and associated
domain definitions can be done faster in an \emph{offline calculation} to
extract relevant geometric information. A systematic method, that allows for
such a splitting into an \emph{offline} pre-processing and an \emph{online}
computation of an effective interfacial evolution problem, are asymptotic
upscaling/homogenization methods. Here, we state two recent upscaling results
which represent homogenized formulations of the microscopic descriptions
$\pmb{[{\bf A}]}$ and $\pmb{[{\bf B1}]}$--$\pmb{[{\bf B2}]}$ stated in
\reff{IEP} and \reff{perCP}--\reff{MIEP}, respectively.

\medskip

{\bf (A) Upscaled formulation for the interfacial transport problem \reff{IEP}.}
The systematic upscaling based on asymptotic two-scale expansions of the form
$
\phi^\epsilon
	=
	\phi^0({\bf x},{\bf x}/\epsilon,t)
	+\epsilon\phi^1({\bf x},{\bf x}/\epsilon,t)
	+\epsilon^2\phi^2({\bf x},{\bf x}/\epsilon,t)
	+\dots
	\,,
$
have been applied in \cite{Schmuck2012prsa,Schmuck2014aml} to derive the following effective macroscopic formulation
of \reff{IEP},  i.e.,
\bsplitl{
\pmb{[{\bf A}_{\rm eff}]}\qquad
\begin{cases}
\quad
p\phi_t
	=
	{\rm div}\brkts{
		M\hat{\rm D}\nabla\brkts{
			\frac{1}{\lambda}w'(\phi)
			-{\rm div}\brkts{
				\frac{\lambda}{p}\hat{\rm D}\nabla\phi
			}
		}
	}
	\,,
	&\quad\text{in }\Omega
	\,,
\\\qquad
M\hat{\rm D}\nabla\brkts{
			\frac{1}{\lambda}w'(\phi)
			-{\rm div}\brkts{
				\frac{\lambda}{p}\hat{\rm D}\nabla\phi
			}
	}\cdot{\bf n}
	= J_l:={\bf n}\cdot{\bf J}_l
	\,,
	&\quad\text{on }\Gamma_l
	\,,
\\\qquad
M\hat{\rm D}\nabla\brkts{
			\frac{1}{\lambda}w'(\phi)
			-{\rm div}\brkts{
				\frac{\lambda}{p}\hat{\rm D}\nabla\phi
			}
	}\cdot{\bf n}
	= J_r:={\bf n}\cdot{\bf J}_r
	\,,
	&\quad\text{on }\Gamma_r
	\,,
\\\qquad
{\bf n}\cdot\hat{\rm D}\nabla\phi
	= 0
	\,,
	&\quad\text{on }\Gamma_t\cup\Gamma_b
	\,,
\\\qquad
{\bf n}\cdot\hat{\rm D}\nabla
	{\rm div}\brkts{
		\frac{1}{p}\hat{\rm D}\nabla\phi
	}
	= 0
	\,,
	&\quad\text{on }I_{ps}\cup\Gamma_t\cup\Gamma_b
	\,,
\\\qquad
\phi({\bf x},0)
	= \phi_0({\bf x})
	&\quad\text{in }\Omega_p
	\,,
\end{cases}
}{A_eff}
where $p:=\frac{\av{Y_p}}{\av{Y}}$ is the porosity and
the porous media correction tensor $\hat{\rm D}=\brcs{d_{ij}}_{1\leq i,j\leq d}$ is defined by
\bsplitl{
d_{ij}
	:=
	\frac{1}{|Y|}\sum_{k=1}^d\int_{Y_p}\brkts{
		\delta_{ij}
		-\delta_{ik}\frac{\partial\xi^j}{\partial y_k}
	}\,d{\bf y}
	\,.
}{pmD}
Finally, the porous media corrector $\xi^k$, $1\leq k\leq d$, solves the following
\emph{reference cell problem}
\bsplitl{
\begin{cases}
\quad
-\sum_{i,j=1}^d\frac{\partial}{\partial y_i}\brkts{
		\delta_{ij}\frac{\partial ({\bf y}_k-\xi^k)}{\partial y_j}
	}
	=
	0\,,
\\\quad
\sum_{i,j=1}^d n_i\brkts{
		\delta_{ij}\frac{\partial ({\bf y}_k-\xi^k)}{\partial y_j}
	}
	=
	0\,,
	&\qquad\text{on }\partial Y_p
	\,,
\\\quad
\xi^k({\bf y}) \text{ is $Y_p$-periodic and $\int_Y\xi^k\,d{\bf y}=0$\,,}
\end{cases}
}{ellY}
which is of the same form as the cell problems obtained in the homogenization of elliptic
equations such as the Laplace and Poisson equations, e.g.
\cite{Allaire1992,Chechkin2007}.

This novel effective macroscopic phase field formulation has been recently rigorously justified by a first error quantification in
\cite{Schmuck2017siap}. If we adopt the notation generally
applied in homogenization theory, then one explicitly states the
$\epsilon$-dependence of solutions (i.e., $\phi^\epsilon$) of the microscopic formulation $\pmb{[{\bf A}]}$ and
since the upscaling consists in passing to the limit $\epsilon\to 0$,
one writes $\phi^0$ for the solution of the effective macroscopic
problem $\pmb{[{\bf A}_{\rm eff}]}$. Hence, if the free energy density
$f_r$ is polynomial, then the error variable
%
${\rm E}^\phi_\epsilon
	:= \phi^\epsilon-(\phi_0+\epsilon\phi_1)
	\,, 
$ 
%
%
where
$\phi_1:=-\sum_{k=1}^d \xi^k_\phi({\bf y})\pd{\phi_0}{x_k}({\bf x},t)$,
satisfies for $0\leq t\leq T$ and $0<T<\infty$ the following estimate
\bsplitl{
\N{E_\epsilon^\phi(\cdot,t)}{H^1(\Omega^\epsilon)}
	:=\brkts{\int_{\Omega^\epsilon}\brkts{
		\av{E_\epsilon^\phi({\bf x},t)}^2
		+\av{\nabla E_\epsilon^\phi({\bf x},t)}^2
	}\,d{\bf x}
	}^{1/2}
	& \leq \epsilon^{1/4}C(T,\Omega,m,\kappa,\lambda)
	\,,
}{ThmErEs}
where $C(T,\Omega,m,\kappa,\lambda)$ is a constant independent of $\epsilon$.

We note that the convergence rate ${\cal O}(\epsilon^{1/4})$ arises due to
the classical argument of relying on a smooth truncation in a neighbourhood
of the boundary. A numerical validation of the error bound \reff{ThmErEs} and
recent developments of novel estimation techniques such as
\cite{Pastukhova2013,Suslina2012}, indicate a linear convergence, i.e.,
${\cal O}(\epsilon)$. Hence, we hope that this first rigorous result for
fourth order problems motivates the future refinement towards a sharp error
quantification.

\begin{figure}
\includegraphics[width=5cm]{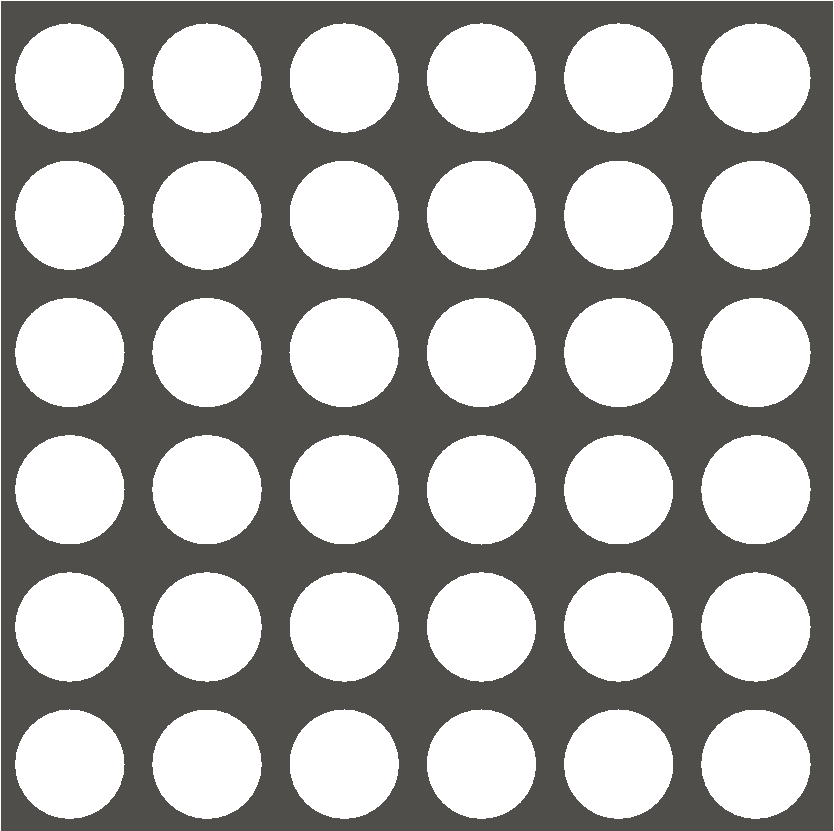}
\hspace{0.2cm}
\includegraphics[width=5cm]{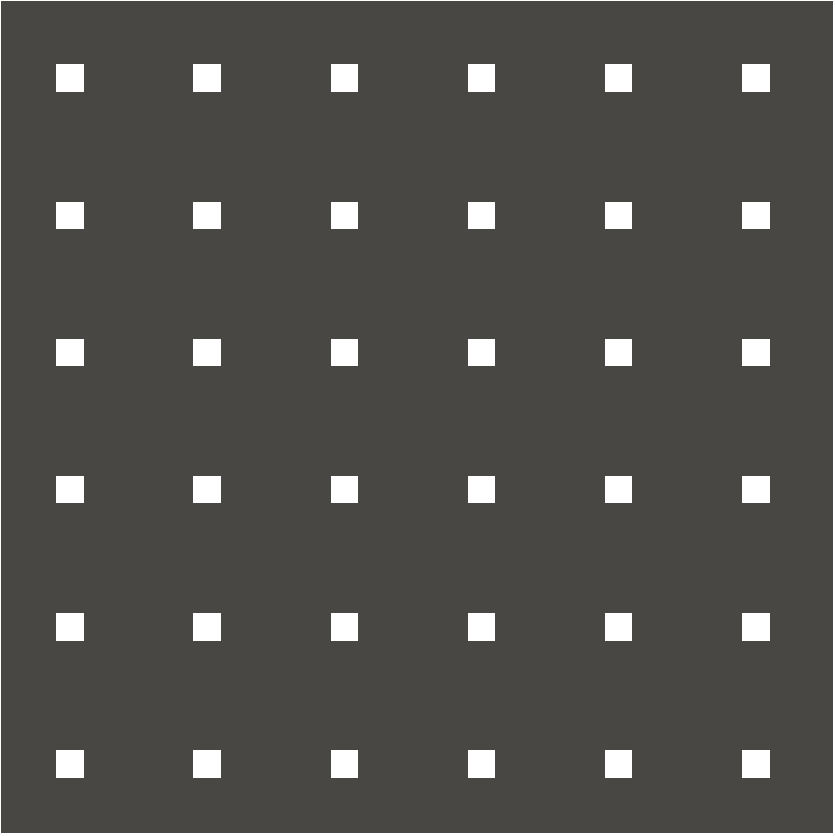}
\hspace{0.2cm}
\includegraphics[width=6.3cm]{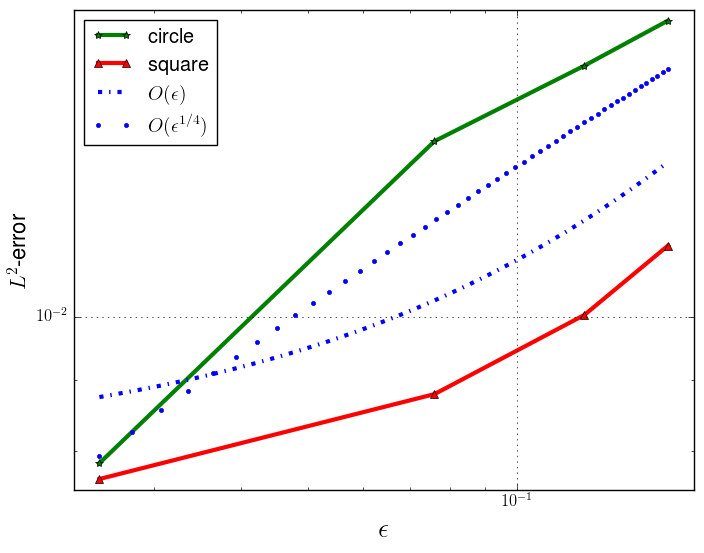}
\caption{The converge for $\epsilon\in\brcs{0.025,0.076,0.125,0.165}$ (right) has been computed
for perforated meshes by removing discs (left) and squares (middle).}
\label{fig:Err}
\end{figure}

\medskip

{\bf (B) Upscaled transport formulation for the quasi-static flow problem \reff{perCP}--\reff{MIEP}.}
Next to systematically and reliably describing interfacial dynamics in strongly heterogeneous systems,
we also want to account for so-called diffusion-dispersion effects of the interface. This latter
phenomenon is well-known for Brownian particles where it has been motivated by the
so-called Taylor-Aris dispersion in \cite{Aris1956,Taylor1953}.
Here, we state the recent upscaling result derived in \cite{Schmuck2013nonl} for the microscopic problem
\reff{perCP}--\reff{MIEP}, i.e., $\pmb{[{\bf B1}]}$--$\pmb{[{\bf B2}]}$,
\bsplitl{
\begin{cases}
\quad
p\phi_t
	=
	{\rm div}\brkts{
			\brkts{
				\hat{\rm C}+\frac{1}{\lambda}\hat{\rm D}w''(\phi)
			}\nabla\phi
	}
	-{\rm div}\brkts{
		\hat{\rm D}\nabla\brkts{
			 \frac{\lambda}{p}{\rm div}\brkts{
			 	\hat{\rm D}\nabla\phi
			}
			-\ol{g}
		}
	}
	\,,
	&\quad\text{in }\Omega
	\,,
\\\qquad
\brkts{
		\brkts{
			\hat{\rm C}+\frac{1}{\lambda}\hat{\rm D}w''(\phi)
		}\nabla\phi
		+
		\nabla\brkts{
			\frac{\lambda}{p}{\rm div}\brkts{
				\hat{\rm D}\nabla\phi
			}
			-\ol{g}
		}
	}\cdot{\bf n}
	= J_l:={\bf n}\cdot{\bf J}_l
	\,,
	&\quad\text{on }\Gamma_l
	\,,
\\\qquad
\brkts{
		\brkts{
			\hat{\rm C}+\frac{1}{\lambda}\hat{\rm D}w''(\phi)
		}\nabla\phi
		+
		\nabla\brkts{
			\frac{\lambda}{p}{\rm div}\brkts{
				\hat{\rm D}\nabla\phi
			}
			-\ol{g}
		}
	}\cdot{\bf n}
	= J_r:={\bf n}\cdot{\bf J}_r
	\,,
	&\quad\text{on }\Gamma_r
	\,,
\\\qquad
{\bf n}\cdot\hat{\rm D}\nabla\phi
	= 0
	\,,
	&\quad\text{on }\Gamma_t\cup\Gamma_b
	\,,
\\\qquad
{\bf n}\cdot\hat{\rm D}\nabla\brkts{
		\frac{\lambda}{p}{\rm div}\brkts{
			\hat{\rm D}\nabla\phi
		}
		-\ol{g}
	}
	= 0
	\,,
	&\quad\text{on }I_{ps}\cup\Gamma_t\cup\Gamma_b
	\,,
\\\qquad
\phi({\bf x},0)
	= \phi_0({\bf x})
	&\quad\text{in }\Omega_p
	\,,
\end{cases}
}{B_eff}
where the porous media correction tensor $\hat{\rm D}$ is defined by \reff{pmD} and \reff{ellY}
as in the case of ${\bf v}={\bf 0}$. At the same time, we have a new
tensor
$\hat{\rm C}:=\brcs{c_{ij}}_{1\leq i,j\leq d}$
contributing to the so-called diffusion-dispersion effects by
\bsplitl{
c_{ik}
	:= \frac{\Pe_0}{\av{Y}}\int_{Y^1}
		\brkts{
			v^i
			-u^i
		}\delta_{ik}\xi^k_{\phi}
	\,d{\bf y}
	\,,
}{c_ik}
with $\xi^k_{\phi}$ being the solution of the cell problem \reff{ellY},
$u^i:=\frac{Pe_0}{\av{Y}}\int_{Y_p} v^i({\bf y})\, d{\bf y}$ for $v^i$
given by \reff{perCP}, and the effective wetting term is given by
$
\ol{g}
	:=
	-\frac{\gamma}{C_h}\frac{1}{\av{Y}}\int_{\partial Y^1}
		a({\bf x},{\bf y})
	\,d{\bf y}
	\,,
$
for wetting characteristics $a$ varying on the macro- and the microscale.

Finally, we emphasize that the advantage of the novel upscaled formulation
\reff{B_eff} is that it allows for a computational decoupling into an
\emph{offline computation} resolving the \emph{microscopic features} of CHeSs
and an \emph{online computation} to solve the low-dimensional, effective
macroscopic phase field equation accounting for diffusion-dispersion
relations. We believe that this novel approach will be useful in many
applications since it allows to take systematic thermodynamic free energies
into account and hence provides a promising framework for investigating
complex reactive multiphase flows.

\subsection{Universal coarsening rates: ${\cal O}(t^{1/3})$-behaviour recovered in heterogeneous media}
\label{sec:CoRa}

The first systematic report on the radial dependence of the coarsening/ripening process
seems to go back to Ostwald in 1900. Since this ``Ostwald ripening'' happens in almost
all phase transition processes and governs the morphology of microstructure,
Ostwald's discovery of this competitive growth phenomenon plays a crucial role in
materials science and related applications.
An important property of the morphology is its self-similarity which one can observe
after sufficiently long coarsening times. The physical explanation for the Ostwald
ripening is that the system tries to minimize its energy by reducing the
system's interfacial area. Moreover, coarsening relies on the fact that a single
large particle has much lower interfacial area than many small particles.
We note that this ripening/coarsening appears in different characteristic length
scales such as distance between particles, particle radius, or the inverse of the
interfacial area per volume, i.e., $L(t) = \av{D}/E_{\rm mix}$, where
$\frac{1}{\av{D}}E_{\rm mix}$
is the volume-averaged interfacial area which relies on the Cahn-Hilliard
free energy density $f_{\rm mix}$, see \reff{E_mix}.
Here, we will focus on this latter length $L$.

About 60 years later since Ostwald's discovery of this growth phenomenon, Lifshitz and
Slyozov \cite{Lifshitz1961} and
Wagner \cite{Wagner1961} proposed a mean field equation whose solution gives
the number of droplets of a particular radius $r$ at time $t$. The following coarsening rate
\bsplitl{
L(t)
	\leq
	C t^{1/3}
	\,,
}{CR}
has been validated experimentally and computationally in \cite{Zhu1999}.
So far, a rigorous proof for \reff{CR} has only been obtained for a time-averaged version in \cite{Kohn2002}, i.e.,
\bsplitl{
\frac{1}{T}\int_0^T
		E_{\rm mix}^2
	\,dt
	\geq
	\frac{C}{T}\int_0^T
		\brkts{t^{-1/3}}^2
	\,dt
	\,.
}{taCR}

\begin{figure}
\includegraphics[width=5.5cm]{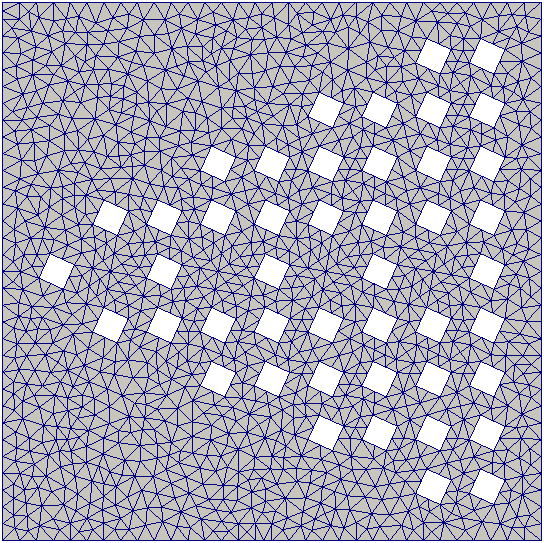}
\includegraphics[width=5.5cm]{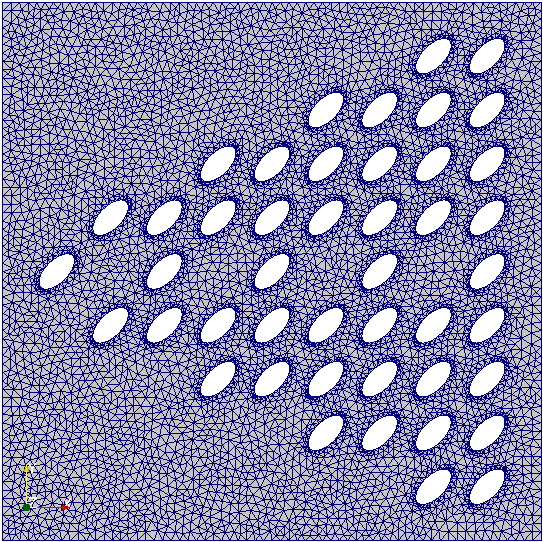}
\includegraphics[width=6.25cm]{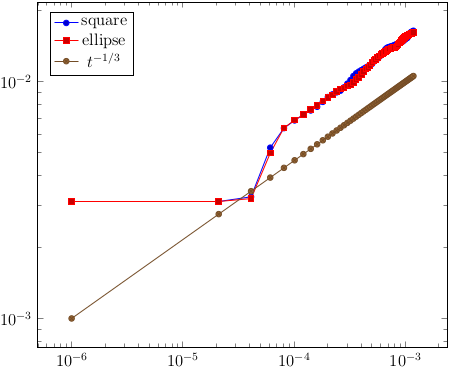}
\caption{{\bf Left and Middle:} Different geometries and porosity gradients. {\bf Right:} Coarsening rate seems to be universal, i.e., independent of pore geometries and porosity gradients.}
\label{fig:CoRa}
\end{figure}

In \cite{Ververis2017}, the authors have recently studied the influence of
heterogeneities such as periodic porous media on the coarsening rate. Here,
we extended this validation towards non-periodic porous media with porosity gradients, see
Fig.~\ref{fig:CoRa}. We observe that the well-known coarsening rate \reff{CR}
for homogeneous media also holds in the context of porous media under neutral
wetting conditions, i.e., a contact angle of $90^{\circ}$. Hence, this indicates
that the exponent $1/3$ in \reff{CR} represents a \emph{universal} coarsening rate.

\section{Application of $\pmb{[{\bf A}_{\rm eff}]}$: upscaled composite cathodes}
\label{sec:appl}

\begin{figure}
\center
\includegraphics[width=8.3cm]{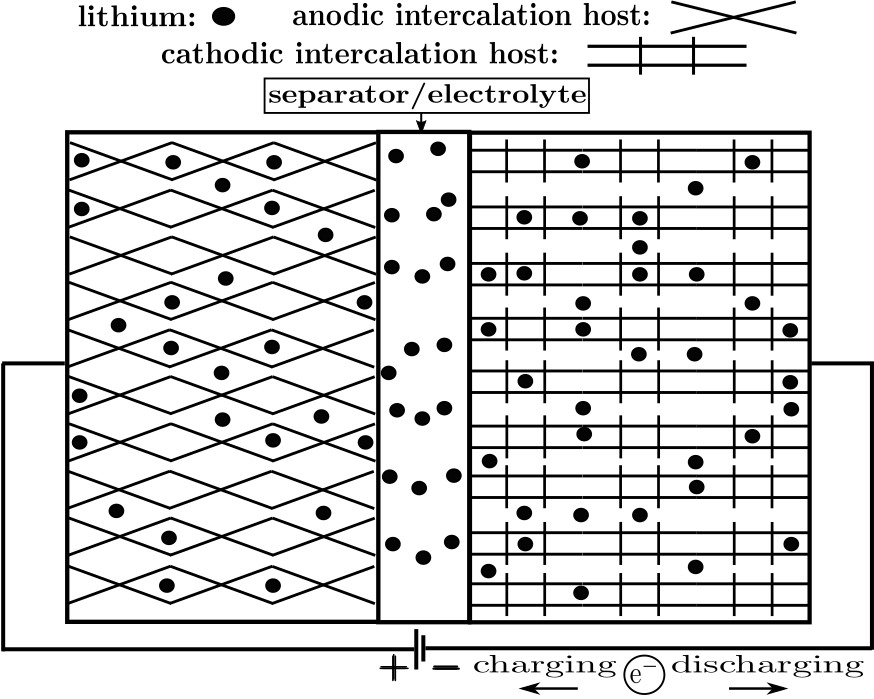}
\hspace{0.5cm}
\includegraphics[width=8.5cm]{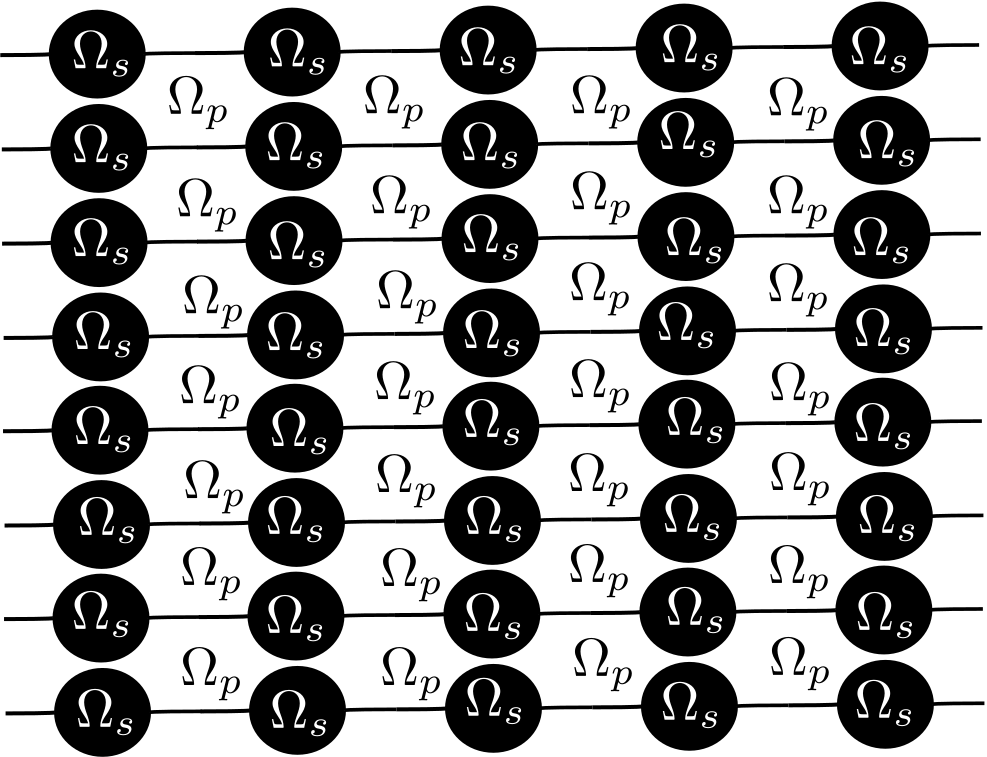}
\caption{{\bf Left:} lithium-ion battery with two active electrodes, i.e.,
anode and cathode separated by a polymer electrolyte. The left- and right-hand vertical boundaries are current collectors. {\bf Right:} A
three-phase composite cathode $\Omega:=\Omega_p\cup\Omega_s\cup\Omega_b$ showing two-phase fiber composites $\Omega_p\cup\Omega_b$ separated vertically by the polymer electrolyte $\Omega_p$. The binder $\Omega_b$ connecting the solid intercalation hosts $\Omega_s$
enables electron conduction towards the current collector.}
\label{fig:2Dcc}
\end{figure}

Batteries represent a delicate CHeSs due to mass and charge transport
through different phases such as an active anode and
a composite cathode which are separated by a polymer electrolyte. Moreover,
the performance of batteries crucially depends on interfacial reactions,
generally described by Butler-Volmer reactions. A schematic design of a lithium-ion battery
is depicted in Fig.~\ref{fig:2Dcc} (left).
For simplicity, we restrict ourselves to the composite cathode which consists
of solid intercalation particles $\Omega_s$, a polymer electrolyte phase
$\Omega_p$, and an electron conducting binder $\Omega_b$ allowing for electron conduction between the solid phase $\Omega_s$. An example of
a composite cathode is given in Fig.~\ref{fig:2Dcc} (right), which
represents horizontal fibers separated by a polymer electrolyte.
We note that an effective model formulation for composite cathodes
has been proposed in \cite{Doyle1993} with the help of a so-called
shrinking core description relying on radial and
\emph{classical diffusion}. In this section, we present a recently derived
generalization towards an effective macroscopic formulation accounting
for interstitial diffusion in heterogeneous domains.

Motivated by the experimental fact that crystalline intercalation hosts
of composite cathodes can phase separate, and by the increased interest in
describing lithium intercalation by a phase field equation as initiated
in \cite{Han2004} and further developed in \cite{Bai2011,Bazant2013accChemRes,Burch2008}, an effective composite
cathode formulation based on phase field driven intercalation and dilute
electrolytes has been systematically derived recently in
\cite{Schmuck2017amrx}. Hence,
the incompressible momentum ${\bf V}$, the densities
$C^+$ and $C^-$
of positively and negatively charged ions, respectively,
the electrostatic potentials $\Psi_s$ and $\Psi_p$
for the solid and electrolyte phase, respectively,
as well as the density of intercalated lithium $C^s$
are described by the following novel
\emph{upscaled composite cathode system},
\bsplitl{
\begin{cases}
\quad
{\bf V}({\bf x},t)
	=
	-\frac{\hat{\kappa}}{\mu}\brkts{
		{\bf f}^0-\nabla P
	}
	&\quad\text{in }\Omega
	\,,
\\
\qquad
	{\rm div}\,{\bf V}
	= 0
	\,,
	&\quad\text{in }\Omega
	\,,
\\
\quad
p\partial_t C^+
	+\Pe{\bf V}\cdot\nabla C^+
	= {\rm div}\brkts{
			\hat{\mathbb{D}}\nabla C^+
			+z_+C^+\hat{\mathbb{M}}\nabla\Psi_p
	}
	-\ol{\beta}_+R(C^+,C^s,\eta^0,U^0)
	&\quad\text{in }\Omega
	\,,
\\
\quad
p\partial_tC^-
	+\Pe {\bf V}\cdot\nabla C^-
	= {\rm div}\brkts{
			\hat{\mathbb{D}}\nabla C^-
			-z_-C^-\hat{\mathbb{M}}\nabla\Psi_p
	}
	&\quad\text{in }\Omega
	\,,
\\
\quad
-{\rm div}\brkts{
		\hat{\varepsilon}\nabla\Psi_p
	}
	= p\brkts{z_+C^++z_-C^-}
	&\quad\text{in }\Omega
	\,,
\\
\quad
-{\rm div}\brkts{
		\hat{\Sigma}\nabla\Psi_s
	}
	= \ol{\beta}_\psi R(C^+,C^s,\eta^0,U^0)
	&\quad\text{in }\Omega
	\,,
\\
\quad
q\partial_tC^s
	= {\rm div}\brkts{
		\hat{\mathbb{M}}_s\brkts{
			\nabla f_r'(C^s)
			+\frac\lambda q{\rm div}\brkts{
				\hat{\mathbb{M}}_s\nabla C^s
			}
		}
	}
	+\ol{\beta}_s R(C^+,C^s,\eta^0,U^0)
	&\quad\text{in }\Omega
	\,,
\end{cases}
}{UCCE}
where
$p=\frac{|Y_p|}{|Y|}$ is the porosity (see Fig.~\ref{fig:refCell}),
$q=1-p$,
${\bf f}^0=-\eta\brkts{C^+-C^-}\hat{\bb M}\nabla\Psi_p$ an
effective Coulomb force, and the material's correction tensors
$\hat{\kappa}=\brcs{\ol{\kappa}_{ij}}_{i,j=1}^d$,
$\hat{\bb M}=\brcs{\ol{m}_{ij}}_{i,j=1}^d$,
$\hat{\bb D}=\brcs{\ol{d}_{ij}}_{i,j=1}^d$,
$\hat{\varepsilon}=\brcs{\ol{\varepsilon}_{ij}}_{i,j=1}^d$,
$\hat{\Sigma}=\brcs{\ol{\sigma}_{ij}}_{i,j=1}^d$,
and
$\hat{\bb M}_s=\brcs{\ol{m}^s_{ij}}_{i,j=1}^d$
are defined by standard cell problems arising in the homogenization
theory (see \reff{ellY} and \reff{pmD} for instance) and therefore we refer the interested reader to
\cite{Schmuck2017amrx}. Finally, the interfacial Butler-Volmer
reactions
\bsplitl{
R
	= i_{ps}
	\ebrkts{
		\frac{C^s}{c_{sm}}\exp\brkts{
			\alpha_{sm}\frac{\alpha_aF}{RT}(\eta^0-U^0)
		}
		-\frac{(c_{\rm sm}-C^s)}{c_{sm}}\exp\brkts{
			-\alpha_{sm}\frac{\alpha_cF}{RT}(\eta^0-U^0)
		}
	}
	\,,
}{ieps}
appear in the upscaled system \reff{UCCE} as bulk
equations and show
the important coupling parameters
$\ol{\beta}_+= \frac{i_0 L |\Lambda|}{eD}$,
$\ol{\beta}_\psi=\frac{F}{RT}\frac{i_{ps}L|\Lambda|}{\sigma_{ref}}$,
$\ol{\beta}_s=\frac{i_0 L |\Lambda|}{e M_s}$,
for $\sigma_{ref}=\frac{\varepsilon_p D}{\lambda_D^2}\,.$
Moreover,
$\eta^0=\Psi_s-\Psi_p$
denotes the overpotential and
$
U^0
$
represents
the open circuit potential. Finally, the variable
$
i_{ps}
	= Fk_{\rm fb}(c_{\rm pm}^{\epsilon_p}-c^{\epsilon_p})^{\alpha_c}(c^{\epsilon_p})^{\alpha_a}
$
is the so-called exchange current density and
$
c_{sm}^{\epsilon_p}
	= {\rm exp}\brkts{\alpha_{sm}}
	\,.
$

The novel system \reff{UCCE} shares with the crucial modelling initiated in
\cite{Doyle1993} the fact that in the effective macroscopic formulation, the
different phases are superimposed or homogenized (referring to the underlying
upscaling strategy). The main novelty and contribution of \reff{UCCE} is the appearance
of the effective phase field equation \reff{UCCE}$_7$ generalizing the radial
diffusion (shrinking core) formulation proposed in \cite{Doyle1993} towards a
thermodynamic formulation taking phase separation during the lithium
intercalation into account.

\section{Conclusions}
\label{sec:Concl}

We have presented recent developments to describe interfacial evolution of
binary mixtures founded on the non-equilibrium thermodynamic structure
provided by the reversible-irreversible couplings, called GENERIC. And we
highlighted the increasing interest in establishing a non-equilibrium
variational principle by generalising the least action principle for
reversible systems to acccount for the right irreversible contributions via a
maximum dissipation principle in Section~\ref{sec:2}.

A major part of this article has then been devoted to demonstrate that
reliable upscaling of phase field equations provides a new and thermodynamic
consistent approach to describe multiphase flow in porous media. In fact, the
novel formulations \reff{A_eff} (without flow) and \reff{B_eff} (with flow)
take the underlying, thermodynamic free energy of fluid mixtures into account
in difference to the classical multiphase extension \reff{gDL} of Darcy's
law. It is noteworthy that Darcy's law represents from a thermodynamic point
of view a reduced momentum balance equation. Moreover, under quasi-static
fluid flow defined on a reference cell in \emph{local thermodynamic
equilibrium}, our upscaled/effective multiphase flow formulation includes the
so-called diffusion-dispersion relations which have been intensively studied
in the context of Brownian motion/Fick's diffusion, e.g.
\cite{Aris1956,Taylor1953}. In fact, the effective macroscopic phase field
formulation \reff{A_eff} has been analytically and computationally validated
by error estimates, i.e., inequality \reff{ThmErEs} and Fig.~\ref{fig:Err}
(right), respectively. Additionally, we investigated the effect of
heterogeneities, e.g. perforated domains with porosity gradients as depicted
in Fig.~\ref{fig:CoRa}, on the coarsening rate and, interestingly, we observe
that the coarsening rate ${\cal O}(t^{1/3})$, well-known for homogeneous
domains, also holds in porous media and hence seems to represent a
\emph{universal property}.

Of course the Cahn-Hilliard phase field equation \cite{Cahn1958} has a long
history going back to 1958. Since then, there is a continuously increasing
interest in applying the mean field formulation in a wide spectrum of fields
including physics, material science, biology, and fluid dynamics to mention
but a few. We believe that the novel multiphase flow/interfacial evolution
equations we outlined, show promise for a wide range of scientific,
engineering, and industrial applications. And we hope that they can motivate
further studies on the use of non-equilibrium thermodynamic framework we
described for problems where heterogeneities play a crucial role. A rather
novel direction is battery science as initiated in \cite{Han2004}, where
the phase field model has been motivated as a reliable description for
interstitial diffusion. This has found increasing interest in
computational material science and electrochemistry and hence motivated us to present here the extension of this
description to systematically account for highly heterogeneous electrodes
such as composite cathodes, see Fig.~\ref{fig:2Dcc} for instance.

\section*{Acknowledgements}
We acknowledge financial support by the Engineering and Physical Sciences
Research Coun- cil of the UK through grants 
EP/H034587/1, 
EP/L027186/1,
EP/L025159/1, 
EP/L020564/1, 
EP/K008595/1, 
EP/P031587/1, 
EP/L024926/1, 
EP/L020564/1, and EP/P011713/1. MS would like to
thank H.C. \"Ottinger (ETH Z\"urich) for the time and discussions of
interdisciplinary physical and mathematical research topics in Spring 2017 as
well as the whole Polymer Physics group at ETH for hospitality. It was during
this visit where MS became aware of A. Jelic's PhD thesis elaborating the
GENERIC aspect of the Cahn-Hilliard equation.


\bibliographystyle{plain} 
\bibliography{pmchcomp} 
          %

\bigskip


\end{document}